\documentclass[a4paper,11pt]{article}
\usepackage{jheppub}
\usepackage[T1]{fontenc}
\usepackage[utf8]{inputenc}
\usepackage{amsmath}
\usepackage{amsfonts}
\usepackage{amssymb}
\usepackage{graphicx}
\usepackage{relsize}
\usepackage{pstricks}
\usepackage{color}
\usepackage{tikz}
\usepackage{booktabs}
\usepackage{csquotes}
\usepackage{blindtext}
\usepackage{imakeidx}
\indexsetup{firstpagestyle=empty}
\makeindex[columns=2,title=Alphabetical Index,intoc]

\newcommand{\eq}{\begin{eqnarray}}
\newcommand{\en}{\end{eqnarray}}
\newcommand{\ra}{\rangle}
\newcommand{\la}{\langle}

\newcommand{\bfk}{{\bf k}_{\perp}}

\newcommand{\seq}{\begin{subequations}}
\newcommand{\sen}{\end{subequations}}

\begin{document}

\title{\boldmath QCD at the amplitude level: Fock state interference
  in heavy quark electroproduction}

\author[a,b,c,1]{Valery E. Lyubovitskij,\note{Corresponding author}}
\author[b]{Ivan Schmidt}

\affiliation[a]{Institut f\"ur Theoretische Physik, Universit\"at T\"ubingen, \\
	Kepler Center for Astro and Particle Physics, \\ 
	Auf der Morgenstelle 14, D-72076 T\"ubingen, Germany}
\affiliation[b]{Departamento de F\'\i sica y Centro Cient\'\i fico
	Tecnol\'ogico de Valpara\'\i so-CCTVal, \\ 
        Universidad T\'ecnica Federico Santa Mar\'\i a, Casilla 110-V,
        Valpara\'\i so, Chile}
\affiliation[c]{Millennium Institute for Subatomic Physics
at the High-Energy Frontier (SAPHIR) of ANID, \\
Fern\'andez Concha 700, Santiago, Chile}

\emailAdd{valeri.lyubovitskij@uni-tuebingen.de}
\emailAdd{ivan.schmidt@usm.cl}

\abstract{
Quantum chromodynamics (QCD) rigorously predicts the existence of  
both nonperturbative intrinsic and perturbative extrinsic heavy quark  
contents of nucleons. In this article we discuss the heavy quark
electroproduction on protons induced by the Fock states $|uud+g\ra$
of three valence quarks in the proton and a nonperturbative gluon,
and the $|uud+Q\bar Q\ra$ non-perturbative state of three valence quarks
and a heavy quark-antiquark pair. The first one gives the perturbative
contribution when the gluon produces a heavy quark pair, while the second
is the intrinsic part, and they produce amplitude interference.
We use nonperturbative light-front wavefunctions for these Fock states, which
are computed using the color-confining light-front holographic QCD theory.
Due to interference of the amplitudes corresponding to the intrinsic
and extrinsic heavy quark contribution to the proton a novel $Q$ vs. $\bar Q$
asymmetry emerges in the differential cross section of the electroproduction
off a proton $d^2\sigma_{e^-p}/(dx dQ^2)$. Our analysis proposed
a novel asymmetry in QCD between intrinsic and extrinsic heavy quark
content in nucleon and provides new insights
into the physics of heavy quark phenomena in QCD
at the amplitude level. This asymmetry is similar to the
Brodsky-Gillespie asymmetry discovered before in QED case
and confirmed at DESY.} 
  

\flushbottom
            
\maketitle

\section{Introduction}

The hard scattering model is the main tool for the theoretical analysis of
the physical processes that happen in hadron colliders.
Its basic ingredients are the parton distribution functions,
the hard partonic scattering cross section and in some cases the parton-hadron 
fragmentation functions. In most cases the collinear approximation is used, in which 
the parton transverse momenta in the distribution functions is ignored. The theoretical 
foundations of this model rely on the factorization theorems, which have been proven 
in specific cases. Nevertheless, although there has been a tremendous amount
of experimental work in abstracting the parton distribution functions from data and
theoretical work in getting the hard scattering cross section from perturbative QCD,
the size of the corrections to the model are not known. All that is known is that these
corrections are going to be suppressed by the hard scale that is involved in
the parton cross section. One correction comes from the fact that the transverse momenta
of the partons in the distribution functions is ignored, but more importantly,
there is a multiplication of probabilities (the hard 
scattering model diagram is not a Feynman diagram), which is certainly an approximation. 
One way to look at this last point is that there are other intermediate states than the 
leading parton-parton scattering, that can participate in the process. For example, 
in $J/\psi$ production, in which the main contribution comes form gluon-gluon scattering, 
there is the contribution in which two gluons from one hadron interacts with one gluon from 
the other~\cite{Schmidt:2018gep}.

From the point of view of obtaining a better knowledge of the properties of QCD, 
it is therefore interesting to look for reactions in which QCD at the amplitude level is 
considered. For example, in nuclear shadowing there is interference of Pomeron and Reggeon 
amplitudes~\cite{Brodsky:2004qa}, and in single spin asymmetries there is interference of
two amplitudes which have different proton spin $J_z = \pm 1/2$ but couple to the same
final-state~\cite{Brodsky:2002cx,Lyubovitskij:2022vcl}. Notice that in 
the hard scattering diagram the blob in which the parton comes out of the proton represents 
a distribution function, but in cases in which amplitudes are considered, it represents 
a light-front wavefunction (LFWF) or Fock state, whose square is related to the parton 
distribution.

One of the rigorous properties of QCD is the existence of intrinsic heavy quarks
in the fundamental structure of hadrons, which is part of the Fock state expansion of the proton.
In fact, the existence of both nonperturbative
intrinsic and perturbative extrinsic heavy quark contents of the nucleon~\cite{Brodsky:1981se},
has now been confirmed by many experiments at world-wide facilities. 
As first noted in~\cite{Brodsky:1981se}, 
the intrinsic contribution comes from the non-perturbative Fock state $|uud+Q\bar Q\ra$. 
For a review see, e.g., Ref.~\cite{Brodsky:2015fna}.  
An even more surprising and novel feature resulting from intrinsic heavy quarks is the strong
asymmetry between the $Q(x)$ and $\bar Q(x)$ distributions in the nucleon eigenstate. 
The $c(x)$ vs. $\bar c(x)$  asymmetry in the proton has recently been demonstrated by a lattice
gauge theory analysis, where the charm quark distribution dominates at large $x$,
and we will use this result in our heavy quark electroproduction analysis, which will
obviously produce a corresponding asymmetry in this process. 
This asymmetry was discovered in Ref.~\cite{Sufian:2020coz} and 
later was confirmed in Ref.~\cite{Brodsky:2022kef}. However, there is an additional
asymmetry describing heavy quark content in the proton. This asymmetry arises
due to interference of intrinsic and extrinsic heavy quark distributions in the proton. 
It was discovered before in case of QED. In QED, the interference of the first-Born and
second-Born amplitudes~\cite{Brodsky:1968rd} 
leads to an $\ell$ vs. $\bar \ell$ asymmetry in the Bethe-Heitler
cross section for lepton pairs produced on nuclei
$\gamma Z \to \ell^+ \ell^- Z$. The lepton asymmetry for electron-positron
pairs was measured at DESY~\cite{Asbury:1967hbp}. 

\begin{figure}[htb]
\begin{center}
\epsfig{figure=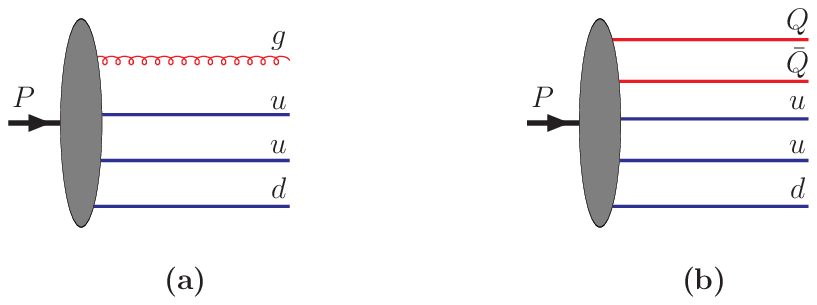,scale=.65}
\caption{Proton Fock states with twist $\tau=4$ and $\tau=5$ 
producing heavy quark-antiquark contributions: 
(a) $|uud+g\ra$ state and (b) $|uud+Q\bar Q\ra$ state. 
\label{fig1}}
\end{center}

\begin{center}
\epsfig{figure=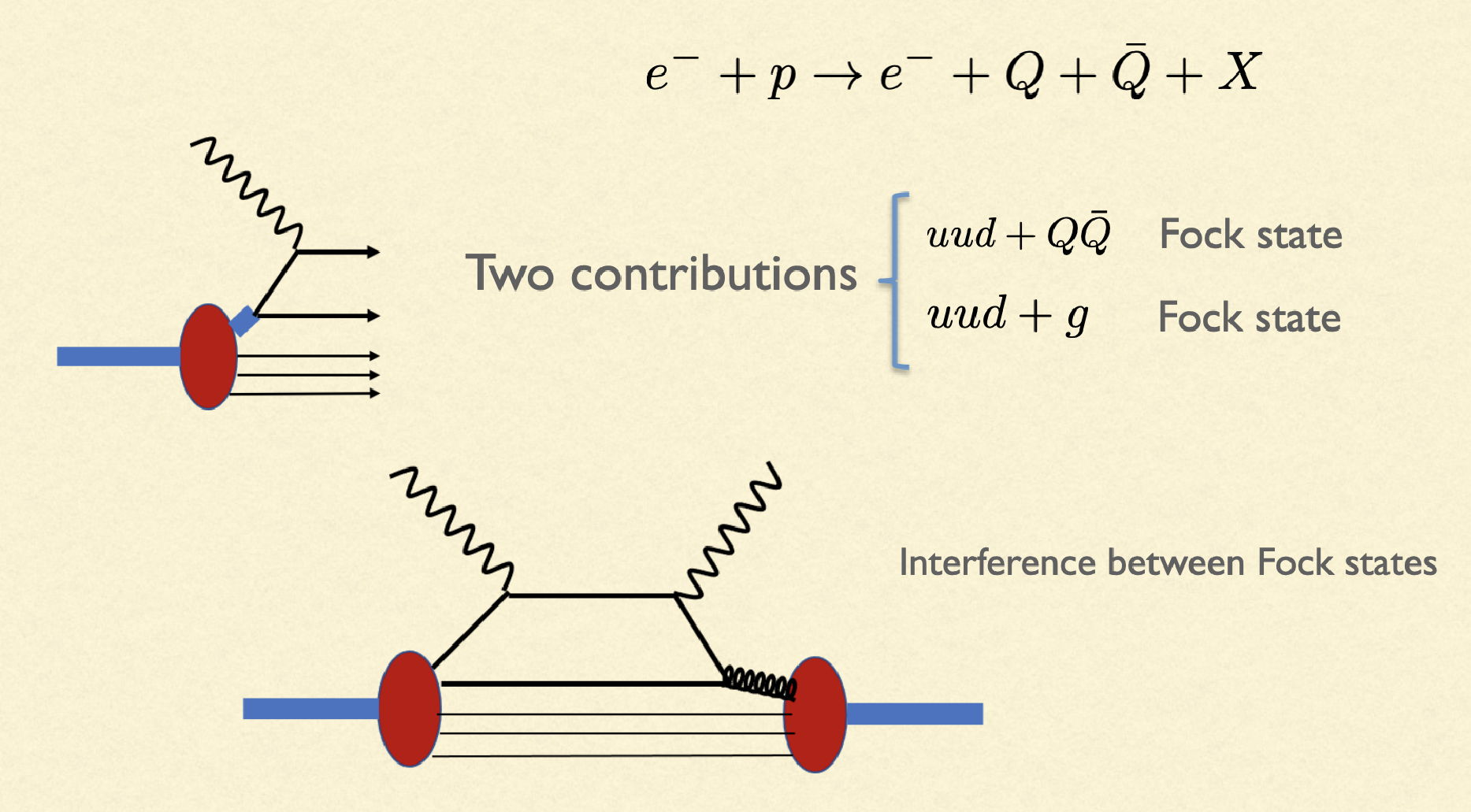,scale=.27}
\caption{Asymmetry in the $e^- + p \to e^- + Q + \bar Q + X$ reaction.
\label{fig2}}
\end{center}
\end{figure}

In particular, in the present paper we will study the analogous effects due to
the interference of QCD amplitudes for the leading Fock states
of the proton producing heavy quarks.
These Fock states are shown in Fig.~\ref{fig1}:
(a) the $|uud+g\ra$ Fock state describing 
the three valence quarks in the proton
and a nonperturbative gluon which produces a heavy quark pair
with odd charge conjugation $C_Q |Q> = - |\bar Q>$ and
(b) the $|uud+Q\bar Q\ra$ state describing the five-quark state,
which is the bound state of the three valence quarks plus
a non-perturbative intrinsic heavy quark-antiquark pair with both even
and odd charge conjugation components.
The interference contributing to
the $e^- + p \to e^- + Q + \bar Q + X$ reaction is given by
the imaginary part of the diagram shown in Fig.~\ref{fig2}.
Our consideration is based on the factorization picture, 
which is valid at large $Q^2$. In comparison with Ref.~\cite{Brodsky:2022kef} 
we consider the interference the Fock states
producing intrinsic and extrinsic heavy quark content in
the nucleon. It leads to a novel asymmetry in QCD, which is similar
to the QED case (as stressed in previous paragraph) and differed from
asymmetry induced by intrinsic heavy quark content in the nucleon
discussed in Refs.~\cite{Sufian:2020coz} and~\cite{Brodsky:2022kef}.

\section{Framework for interference of intrinsic and  extrinsic heavy quarks in the nucleon}

In order to calculate the effects of the interference, we
compute the square of the matrix element corresponding to the 
scattering amplitude $e^- + p \to e^- + Q + \bar{Q} + X$.
The amplitude is given by the sum of the two terms induced by
the two Fock states shown in Fig.~\ref{fig2}.
Therefore, one gets three contributions: two squares of
the amplitudes generated by the individual Fock states
and also their interference.

Following the ideas
of holographic QCD~\cite{Brodsky:2014yha}-\cite{Lyubovitskij:2020xqj} 
we proposed~\cite{Brodsky:2022kef} the heavy quark and heavy antiquark
parton distribution functions (PDFs)
$Q_{\rm in}(x)$ and $\bar Q_{\rm in}(x)$ encoding
intrinsic heavy quark and heavy antiquark content in the nucleon.  
The PDFs $Q_{\rm in}(x)$ and $\bar Q_{\rm in}(x)$ are expressed in terms of
the LFWFs $\psi^{{\rm in}; \lambda_N}_{Q; \lambda_Q, L_z}(x,\bfk)$  
and $\psi^{{\rm in}; \lambda_N}_{\bar Q; \lambda_{\bar Q}, L_z}(x,\bfk)$
with struck heavy quark $Q$ and heavy antiquark $\bar Q$, respectively: 
\eq\label{Qin_barQin} 
Q_{\rm in}(x) &=& 
\int \frac{d^2\bfk}{16\pi^3} \,
\biggl[ |\psi^{{\rm in}; \uparrow}_{Q; + \frac{1}{2}, 0}(x,\bfk)|^2
      + |\psi^{{\rm in}; \uparrow}_{Q;- \frac{1}{2}, +1}(x,\bfk)|^2
      \biggr] 
\,, \nonumber\\
\bar Q_{\rm in}(x) &=& 
\int \frac{d^2\bfk}{16\pi^3} \,
\biggl[ |\psi^{{\rm in}; \uparrow}_{\bar Q; + \frac{1}{2}, 0}(x,\bfk)|^2
      + |\psi^{{\rm in}; \uparrow}_{\bar Q;- \frac{1}{2}, +1}(x,\bfk)|^2
\biggr] \,.
\en
The LFWFs introduced in Eq.~(\ref{Qin_barQin}) correspond  
to the Fock state $|uud+Q \bar Q\ra$ with specific quantum numbers ---
helicity for the proton $\lambda_N = \uparrow, \downarrow$,
helicity for the struck heavy quark or heavy antiquark
$\lambda_Q, \lambda_{\bar Q} = + \frac{1}{2}, -\frac{1}{2}$, 
and $z$ projection of the angular orbital momentum~$L_z$.

For the heavy quark PDF $Q_{\rm in}(x)$ 
the result obtained in LF QCD~\cite{Brodsky:1981se} reads  
\eq
Q_{\rm in}(x) = N_{Q_{\rm in}} \, x^2 \, \Big[ (1-x) (1 + 10 x + x^2)
  + 6 \, x \, (1+x) \log(x) \Big] \,.
\en 
Note that $Q_{\rm in}(x)$ scales at large $x \to 1$ as $(1-x)^5$. 
Here $N_{Q_{\rm in}}$ is the normalization constant fixed from data.
For example,  as pointed out in Ref.~\cite{Brodsky:2015fna},
in the case of the charm distribution $N_{c_{\rm in}} = 6$,
if there is a 1\% intrinsic charm contribution to the proton PDF.
Such a choice was motivated by an estimate of the magnitude
of the diffractive production of the $\Lambda_c$ baryon in the
$pp \to p \Lambda_c X$ reaction~\cite{Brodsky:1981se}, which is consistent
with the MIT bag-model estimate~\cite{Donoghue:1977qp} of the probability
of finding a five-quark $|uud+Q\bar Q\ra$ configuration the nucleon
at the order of 1\%. In the case of the bottom distribution, 
$N_{b_{\rm in}} = 6 \, (m_c/m_b)^2$~\cite{Brodsky:2015fna}. 
Heavy antiquark PDF $\bar Q_{\rm in}(x)$ is related to $Q_{\rm in}(x)$ as
$\bar Q_{\rm in}(x) = 7/5 (1-x) \, Q_{\rm in}(x)$~\cite{Brodsky:2022kef}.
This form of the heavy antiquark PDF was motivated
by Ref.~\cite{Sufian:2020coz} and the constraints that the zero moments
of the $Q_{\rm in}(x)$ and $\bar Q_{\rm in}(x)$ PDFs should be equal, 
or that the zero moment of the asymmetry
$Q_{\rm as}(x) = Q_{\rm in}(x) - \bar Q_{\rm in}(x)$
should vanish. Our prediction for the first moment of the heavy quark
asymmetry $\la x \ra_{c-\bar c} = \int_0^1 dx \, x \, Q_{\rm as}(x)
= 1/3500 \simeq 0.00029$~\cite{Brodsky:2022kef} was in order of
magnitude  agreement with the prediction 
of Ref.~\cite{Sufian:2020coz} $\la x \ra_{c-\bar c} = 0.00047(15)$.
Also our prediction for the fraction of the proton momentum
$[Q]  =  \int_0^1 dx \, x \, \Big[Q_{\rm in}(x) + \bar Q_{\rm in}(x)\Big] = 0.54\%$
carried by charm quark and antiquark was in good agreement
with the central value $0.62\%$ of the results recently extracted
by the NNPDF Collaboration~\cite{Ball:2022qks}.   

To fix the explicit form of the LFWFs
$\psi^{{\rm in}; \lambda_N}_{Q; \lambda_Q, L_z}(x,\bfk)$ 
and $\psi^{{\rm in}; \lambda_N}_{\bar Q; \lambda_{\bar Q}, L_z}(x,\bfk)$ 
we propose that they are proportional to
the scalar functions $\varphi_{Q_{\rm in}}(x,\bfk^2)$
and $\varphi_{\bar Q_{\rm in}}(x,\bfk^2)$, which are parametrized   
in terms of the PDFs $Q_{\rm in}(x)$ and $\bar Q_{\rm in}(x)$
of the intrinsic heavy quark and antiquark,
respectively~\cite{Brodsky:2022kef} 
\eq
\varphi_{Q_{\rm in}}(x,\bfk^2) &=& \dfrac{2 \pi \sqrt{2}}{\kappa} \, 
\sqrt{Q_{\rm in}(x)} \, \exp\Big[-\dfrac{\bfk^2}{2\kappa^2}\Big]\,,
\nonumber\\
\varphi_{\bar Q_{\rm in}}(x,\bfk^2) &=& \dfrac{2 \pi \sqrt{2}}{\kappa} \, 
\sqrt{\bar Q_{\rm in}(x)} \, \exp\Big[-\dfrac{\bfk^2}{2\kappa^2}\Big] \,. 
\en
Here $\kappa = 500$ MeV is the dilaton scale. 
The conjecture
$\varphi_{Q_{\rm in}}(x,\bfk^2) \neq \varphi_{\bar Q_{\rm in}}(x,\bfk^2)$
proposed in Ref.~\cite{Brodsky:2022kef}
led to the asymmetry $Q(x)-\bar Q(x)$. 

Finally, the LFWFs $\psi^{{\rm in}; \lambda_N}_{Q; \lambda_Q, L_z}(x,\bfk)$ 
and $\psi^{{\rm in}; \lambda_N}_{\bar Q; \lambda_{\bar Q}, L_z}(x,\bfk)$ read 
\eq
\psi^{{\rm in}; \uparrow}_{Q/\bar Q; + \frac{1}{2}, 0}(x,\bfk) &=&
\psi^{{\rm in}; \downarrow}_{Q/\bar Q; - \frac{1}{2}, 0}(x,\bfk) =
\varphi_{Q_{\rm in}/\bar Q_{\rm in}}(x,\bfk^2) \,, \nonumber\\
\psi^{{\rm in}; \uparrow}_{Q/\bar Q; - \frac{1}{2}, +1}(x,\bfk) &=& 
- \Big[\psi^{{\rm in}; \downarrow}_{Q/\bar Q; + \frac{1}{2}, -1}(x,\bfk)\Big]^\dagger
= \mp \dfrac{k^1 + i k^2}{\kappa} \, \varphi_{Q_{\rm in}/\bar Q_{\rm in}}(x,\bfk^2)
\,,
\en

The LFWFs $\psi^{\lambda_N}_{\lambda_g, \lambda_X}(x,\bfk)$, describing the
Fock state of a gluon $(g)$ and a three-quark spectator $X=(uud)$, 
where $\lambda_g = \pm 1$, and $\lambda_X = \pm \frac{1}{2}$
are the helicities gluon and three-quark spectator,
are listed as~\cite{Lyubovitskij:2020xqj}: 
\eq
\psi^{\uparrow}_{+1, +\frac{1}{2}}(x,\bfk) &=& 
- \Big[\psi^{\downarrow}_{-1, -\frac{1}{2}}(x,\bfk)\Big]^\dagger =
\frac{k^1-ik^2}{\kappa} \, \varphi^{(2)}(x,\bfk^2) \,, 
\nonumber\\
\psi^{\uparrow}_{+1, -\frac{1}{2}}(x,\bfk) &=& 
+ \Big[\psi^{\downarrow}_{-1, +\frac{1}{2}}(x,\bfk)\Big]^\dagger = 
\varphi^{(1)}(x,\bfk^2)\,, \nonumber\\
\psi^{\uparrow}_{-1, +\frac{1}{2}}(x,\bfk) &=& 
- \Big[\psi^{\downarrow}_{+1, -\frac{1}{2}}(x,\bfk)\Big]^\dagger = 
- \frac{k^1+ik^2}{\kappa} \, (1-x) \, \varphi^{(2)}(x,\bfk^2) \,,
\en
where the functions $\varphi^{(1)}(x,\bfk^2)$ and 
$\varphi^{(2)}(x,\bfk^2)$ are expressed through the 
gluon PDF functions $G^\pm(x)$ as~\cite{Lyubovitskij:2020xqj}  
\eq\label{LFWFs_symmetric}
\varphi^{(1)}(x,\bfk^2) &=& 
\frac{4 \pi}{\kappa} \, \sqrt{G^+(x)-\frac{G^-(x)}{(1-x)^2}} \, 
\exp\biggl[- \frac{\bfk^2}{2 \kappa^2} \biggr] \,,
\nonumber\\
\varphi^{(2)}(x,\bfk^2) &=& \frac{4 \pi}{\kappa} \, 
\frac{\sqrt{G^-(x)}}{1-x} \,
\exp\biggl[- \frac{\bfk^2}{2 \kappa^2} \biggr] \,.
\en
Here $G^+(x) = G_{g\uparrow/N\uparrow}(x) = G_{g\downarrow/N\downarrow }(x)$
and  $G^-(x) = G_{g\downarrow/N\uparrow}(x) = G_{g\uparrow/N\downarrow }(x)$
are the helicity-aligned and helicity-anti-aligned  gluon distributions,
respectively, whose combinations define the gluon unpolarized
$G(x) = G^+(x) + G^-(x)$ and polarized $\Delta G(x) = G^+(x) - G^-(x)$ PDFs. 

$G(x)$ and $\Delta G(x)$ are expressed in terms of derived LFWFs
$\psi^{\lambda_N}_{\lambda_g, \lambda_X}(x,\bfk)$ as
\eq
\left(
\begin{array}{r}
G(x)         \\
\Delta  G(x) \\
\end{array}
\right)
= \int \frac{d^2\bfk}{16\pi^3} \,
\biggl[ |\psi^{\uparrow}_{+1, +\frac{1}{2}}(x,\bfk)|^2
+ |\psi^{\uparrow}_{+1, -\frac{1}{2}}(x,\bfk)|^2
\pm  |\psi^{\uparrow}_{-1, +\frac{1}{2}}(x,\bfk)|^2 \biggr] \,. 
\en
For $G^+$ and $G^-$ we use the results that were proposed
in Ref.~\cite{Brodsky:1989db}: 
$G^+(x) = N_g \, (1-x)^4 \, (1 + 4x)/x$, 
$G^-(x) = N_g \, (1-x)^6/x$, 
where $N_g = 0.8967$ is the normalization constant, fixed 
from the first moment of the gluon PDF~\cite{Brodsky:1989db}: 
$\la x_g \ra = \int_0^1 dx x G(x) = (10/21) N_g$. 
For $\la x_g \ra$ we take the central value of the lattice result
$\la x_g \ra = 0.427$~\cite{Alexandrou:2020sml}. 
These densities obey very important model-independent constraints:
(1) at large $x$ the power scaling $G^+(x) \sim (1-x)^4$ and
$G^-(x) \sim (1-x)^6$~\cite{Brodsky:1989db}, which is consistent
with QCD constraints~\cite{Bjorken:1969mm}  
dictated by matching the signs of the quark and gluon helicities and 
the even power scaling of gluon PDFs;
(2) at small $x$ the gluon asymmetry ratio $\Delta G/G$ behaves as 
$\Delta G(x)/G(x) \to 3 x$ and is consistent with Reggeon exchange
arguments~\cite{Brodsky:1988ip}.

The $|uud+g\ra$ state generates the effective LFWFs
$\psi^{{\rm ex}; \lambda_N}_{Q/\bar Q;\lambda_{Q/\bar Q}, L_z}(x,\bfk)$ encoding
the extrinsic heavy quark contribution in the proton, which are
described by the product of $\psi^{\lambda_N}_{\lambda_g, \lambda_X}(x,\bfk)$,
gluon propagator, and the splitting function for the 
annihilation of the gluon into the heavy
quark-antiquark pair 
\eq
\psi^{\rm ex; \uparrow}_{Q/\bar Q; +\frac{1}{2}, -1}(x,\bfk) &=&
- \Big[\psi^{\rm ex; \downarrow}_{Q/\bar Q; -\frac{1}{2}, +1}(x,\bfk)\Big]^\dagger
\ = 
\sqrt{N_{Q_{\rm ex}}} \, \alpha_s \, C_F \, 
\frac{k^1-ik^2}{\kappa} \,  \xi(x) \, \varphi^{(2)}(x,\bfk^2) 
\,, \nonumber\\
\psi^{\rm ex; \uparrow}_{Q/\bar Q; +\frac{1}{2}, 0}(x,\bfk) &=& 
+ \Big[\psi^{\rm ex; \downarrow}_{Q/\bar Q; -\frac{1}{2}, 0}(x,\bfk)\Big]^\dagger = 
\pm \ \sqrt{N_{Q_{\rm ex}}} \, \alpha_s \, C_F \, \xi(x) \, \varphi^{(1)}(x,\bfk^2)  
\,, \nonumber\\
\psi^{\rm ex; \uparrow}_{Q/\bar Q; -\frac{1}{2}, +1}(x,\bfk) &=& 
- \Big[\psi^{\rm ex; \downarrow}_{Q/\bar Q; +\frac{1}{2}, -1}(x,\bfk)\Big]^\dagger
\nonumber\\ 
&=& - \sqrt{N_{Q_{\rm ex}}} \, \alpha_s \, C_F \, 
\frac{k^1+ik^2}{\kappa} \, \xi(x) \, (1-x) \, \varphi^{(2)}(x,\bfk^2)  
\,,
\en
where $\alpha_s$ is the strong coupling, 
$C_F = 4/3$ is color summation factor,
$\xi(x) = x (1-x) [x^2 + (1-x)^2]$, 
and $N_{Q_{\rm ex}}$ is the normalization constant
of the extrinsic heavy quark contribution, 
which is fixed from data. Note that the $\alpha_s$ 
at the one-loop is calculated according the formula 
$\alpha_s = \alpha_s(\mu^2) = 4 \pi/(\beta_0 \log(\mu^2/\Lambda_{\rm QCD}^2))$,
where $\mu = 2 m_Q$ is the scale taken at the value of two heavy quark masses,
$\Lambda_{\rm QCD} = 0.226$ GeV is the QCD scale parameter used in 
our previous paper~\cite{Lyubovitskij:2022vcl}, 
$\beta_0 = (11/3) N_c - (2/3) N_f$ is the leading coefficient in the
$\alpha_s$ expansion of the QCD $\beta$ function. In particular,
for the charm case we have $N_f = 4$, $\mu \simeq 3$~GeV,
and therefore $\alpha_s = 0.3$. 

The extrinsic PDFs $Q_{\rm ex}(x)$ and $\bar Q_{\rm ex}(x)$
are degenerate $Q_{\rm ex}(x) = \bar Q_{\rm ex}(x)$ 
and are expressed in terms of derived LFWFs as 
\eq 
Q_{\rm ex}(x) &=& \bar Q_{\rm ex}(x) \ = \
\int \frac{d^2\bfk}{16\pi^3} \,
\biggl[ |\psi^{{\rm ex}; \uparrow}_{Q/\bar Q; +\frac{1}{2},  0}(x,\bfk)|^2
      + |\psi^{{\rm ex}; \uparrow}_{Q/\bar Q; -\frac{1}{2}, +1}(x,\bfk)|^2
      \nonumber\\
      &+& |\psi^{{\rm ex}; \uparrow}_{Q/\bar Q; +\frac{1}{2}, -1}(x,\bfk)|^2
      \biggr] 
= N_{Q_{\rm ex}} \, \alpha_s^2 \, C_F^2 \, \xi^2(x) \, G(x)
\,.
\en

\section{Electroproduction of heavy quarks and a novel asymmetry in QCD}

Next, we specify the matrix element describing the elastic
$e^- +  Q/\bar Q \to e^- +  Q/\bar Q$ scattering. It is given by 
\eq 
{\cal M}(e^- + Q/\bar Q \to e^- + Q/\bar Q) &=&
\frac{4 \pi}{t} \, \alpha e_{Q/\bar Q} \, 
\bar u_e(p_3,s_3) \gamma_\mu u_e(p_1,s_1)
\nonumber\\
&\times&
\bar u_{Q/\bar Q}(p_4,s_4) \gamma^\mu u_{Q/\bar Q}(p_2,s_2)
\,,
\en 
where $\alpha = 1/137.036$ is the fine-structure coupling; 
$e_{Q/\bar Q}$ is electric charge of heavy quark/antiquark; 
$u_e(p,s)$, $u_{Q/\bar Q}(p,s)$ are the spinors of electron 
and heavy quark (antiquark), which will be taken to be on-shell.
Neglecting the masses of electron and heavy quark,  
the square of the amplitude ${\cal M}(e^- + Q/\bar Q \to e^- + Q/\bar Q)$
summed over polarizations of the fermions is equal to 
$\overline{|{\cal M}|^2} = 128 \, (\pi\alpha e_Q)^2 \ (s^2 + u^2)/t^2$
for both quark and antiquark case. Note that our formalism is correct
for any $Q^2$, although as an example, we apply it for $Q^2 \gg m^2$
in the present paper. Taking into account the effects of finite
heavy quark masses $m$ will be done in future work. 
We use the Mandelstam partonic level variables $s, t, u$
in the proton rest frame, with 
$s = (p_1+p_2)^2 = 2 x E_{\ell} m_N $,
$t = (p_1-p_3)^2 = q^2 = -Q^2 = - s y$,
$u = (p_1-p_4)^2 = - s (1-y)$, and
$s + t + u = 0$,
where $x = p_2/P = Q^2/(2 P \cdot q)$ is the Bjorken variable, 
$y = (q \cdot P)/(p_1 \cdot P)$ is the rapidity, 
$P$ and $m_N$ are the proton momentum and mass. 

Now we calculate the square of the amplitude
$M(e^- + p \to e^- + Q + \bar Q + X)$ induced
by the two Fock states $|uud+Q\bar Q\ra$ and $|uud+g\ra$ 
which is averaged over the spins of initial electron
and proton and summed over the final fermions polarizations: 
\eq\label{fullM2}
|M^{ep}(x,\bfk^2)|^2 = \frac{1}{4} \, \sum\limits_{\rm pol}
|M(e^- + p \to e^- + Q + \bar Q + X)|^2 =
f(x,\bfk^2) \ \frac{\overline{|{\cal M}|^2}}{4} 
\en 
which is the product of the effective transverse-momentum distribution (TMD)
function $f(x,\bfk^2)$ and $\overline{|{\cal M}|^2}/4$.  
The TMD $f(x,\bfk^2) = \sum_{i=1}^3 f_i(x,\bfk^2)$ is split 
into diagonal contributions of intrinsic ($f_1$) and extrinsic ($f_2$)
amplitudes, and their interference ($f_3$): 
\eq
f_1(x,\bfk^2)  &=& \biggl[
  \Big(\varphi_{Q_{\rm in}}(x,\bfk^2)\Big)^2
+ \Big(\varphi_{\bar Q_{\rm in}}(x,\bfk^2)\Big)^2 \biggr]
\ \ \Big[1+\dfrac{\bfk^2}{\kappa^2}\Big]
\,, \nonumber\\
f_2(x,\bfk^2) &=&
2  \, N_{Q_{\rm ex}} \, \alpha_s^2 \, C_F^2 \, \xi^2(x) \, 
\Big[\Big(\varphi^{(1)}(x,\bfk^2)\Big)^2
  +  \dfrac{\bfk^2}{\kappa^2} \Big(1 +  (1-x)^2\Big)
\, \Big(\varphi^{(2)}(x,\bfk^2)\Big)^2\Big]
\,, \nonumber\\
f_3(x,\bfk^2) &=& 
2 \sqrt{N_{Q_{\rm ex}}} \, \alpha_s \, C_F \, \xi(x) \, 
\Big[\varphi_{Q_{\rm in}}(x,\bfk^2)   
  - \varphi_{\bar Q_{\rm in}}(x,\bfk^2)\Big]
\nonumber\\
&\times&
\Big[\varphi^{(1)}(x,\bfk^2)
  +  \dfrac{\bfk^2}{\kappa^2} \, (1-x) \, 
\varphi^{(2)}(x,\bfk^2)\Big]
\,.
\en
The differential cross section
for the electroproduction of heavy quarks is given by
\eq
\dfrac{d^2\sigma_{e^-p}}{dx dQ^2} =
\dfrac{1}{16\pi s^2} \, \int \dfrac{d^2\bfk}{16\pi^3} \,
|M^{ep}(x,\bfk^2)|^2  
= \dfrac{2 \pi \alpha^2}{x Q^4} \,
F_2^Q(x) \, \Big(1+ (1-y)^2\Big) \,.
\en
Here $F_2^Q(x)$ is the heavy quark structure function, predicted by
our approach, which includes contributions of both $|uud+Q\bar Q\ra$ 
and $|uud+g\ra$ Fock states and their interference:
\eq
F_2^Q(x) &=& e_Q^2 \, x \, \int \dfrac{d^2\bfk}{16\pi^3} \,
f(x,\bfk^2)  
=  e_Q^2 \, x \, \Big[Q_{\rm full}(x) + \bar Q_{\rm full}(x)\Big]
\,,
\en
where $Q_{\rm full}(x)$ and $\bar Q_{\rm full}(x)$ are the full results
for the heavy quark and heavy antiquark PDFs including intrinsic and
extrinsic contributions, and their interference:
\eq\label{Q_barQ_full}
Q_{\rm full}(x) &=& Q_{\rm in}(x) + Q_{\rm ex}(x)
+ \sqrt{Q_{\rm in}(x)} \, H(x)
\,, \nonumber \\
\bar Q_{\rm full}(x) &=& \bar Q_{\rm in}(x) + Q_{\rm ex}(x)
- \sqrt{\bar Q_{\rm in}(x)} \, H(x)
\,, 
\en
where
\eq
H(x) = \sqrt{2 N_{Q_{\rm ex}}}  \, \alpha_s \, C_F \, \xi(x) 
\ \biggl[\!\sqrt{G^+(x) - \frac{G^-(x)}{(1-x)^2}} 
+ \sqrt{G^-(x)}\biggr]
\en 
Combining the interference terms containing in
$Q_{\rm full}(x)$ and $\bar Q_{\rm full}(x)$ we derive a novel
quantity $\Delta_Q(x)$, which we name asymmetry in the
differential cross section of electroproduction off a proton:
\eq
\Delta_Q(x) &=& H(x) \,
\Big[\sqrt{Q_{\rm in}(x)} - \sqrt{\bar Q_{\rm in}(x)}\,\Big]
\,. 
\en 
We stress again that this asymmetry emerges due to interference of the
amplitudes produced by $|uud+Q\bar Q\ra$ and $|uud+g\ra$ 
Fock states or due the intrinsic-extrinsic heavy quark
interference. This novel phenomena is similar to
the Brodsky-Gillespie asymmetry discovered before
in QED case~\cite{Brodsky:1968rd} 
and confirmed at DESY~\cite{Asbury:1967hbp}.

For convenience one can define the asymmetry in the normalized form
\eq
{\cal P }_{\rm as}(x) = \frac{\Delta_Q(x)}{Q_{\rm in}(x) + \bar Q_{\rm in}(x)}
=  \frac{\sqrt{Q_{\rm in}(x)} - \sqrt{\bar Q_{\rm in}(x)}}
{Q_{\rm in}(x) + \bar Q_{\rm in}(x)} \, H(x) \,. 
\en 

Indeed this asymmetry is induced by the intrinsic-extrinsic heavy quark
interference and this phenomena is similar to the Brodsky-Gillespie asymmetry
discovered before in QED case~\cite{Brodsky:1968rd} 
and confirmed at DESY~\cite{Asbury:1967hbp}. 

\begin{figure}
  \begin{center}
  \epsfig{figure=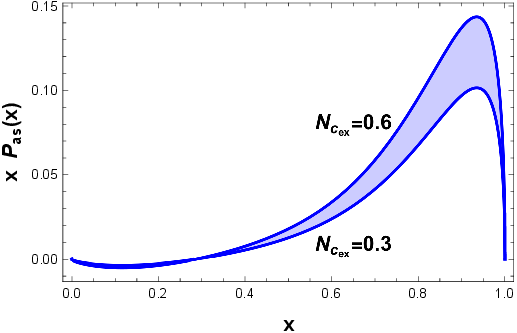,scale=0.75}
  \caption{Asymmetry $x \, {\cal P}_{\rm as}(x)$
  for $N_{c_{\rm ex}}$ varied from 0.3 to 0.6.
\label{fig3}}
\end{center}
\begin{center}
\epsfig{figure=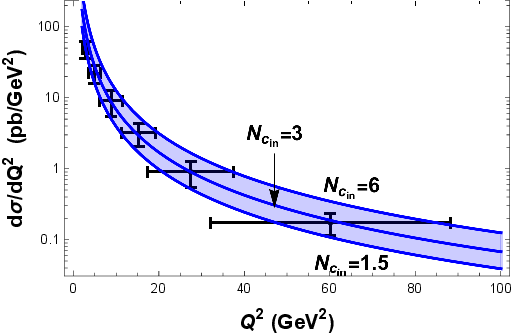,scale=0.75}
\caption{Differential cross section $d\sigma_{e^-p}/dQ^2$ in
comparison with data from H1 Collaboration~\cite{H1:2002voc}.
\label{fig4}}
\end{center}
\end{figure}

Now we consider numerical applications of our findings. 
In Fig.~\ref{fig3} we plot the asymmetry
$x \, {\cal P}_{\rm as}(x)$ as function of $x$
when the extrinsic normalization constant 
$N_{\rm ex}$ is varied from 0.3 to 0.6 and
$N_{c_{\rm in}} = 6$. 
In Fig.~\ref{fig4} we plot our predictions for the
differential cross section $d\sigma_{e^-p}/dQ^2$
at $y=1$ including all charm quark and
antiquark contributions (intrinsic, extrinsic, and
their interference) and compare it with data extracted
from results of the H1 Collaboration at DESY~\cite{H1:2002voc}.
Here we use fixed value of $N_{c_{\rm ex}}=0.3$.  
For completeness we vary the $N_{c_{\rm in}}$ from 1.5 to 6, which
corresponds to the 0.25\%$-$1\% of intrinsic charm in the proton. 
The variation of the $N_{c_{\rm in}}$ is shown by shaded band.
One can see that our curve for the cental value of the $N_{c_{\rm in}}=3$
fits perfectly the data. 

\section{Conclusion}

In conclusion, we have confirmed the existence of a novel asymmetry
${\cal P}_{\rm as}(x)$ in the differential cross section
$d^2\sigma_{e^-p}/(dx dQ^2)$ of the heavy quark pair electroproduction
in the proton. This asymmetry occurs due to interference of
the amplitudes corresponding to the nonperturbative
Fock states $|uud+Q\bar Q\ra$ and $|uud+g\ra$
in the process of the heavy quark pair electroproduction in the proton. 
We presented the study of this asymmetry in QCD at the amplitude level. 
Due to our conjecture that the distributions of heavy quark and antiquark
in the $|uud+Q\bar Q\ra$ are described by different scalar functions 
$\varphi_{Q_{\rm in}}(x,\bfk^2)$ $\ne$ $\varphi_{\bar Q_{\rm in}}(x,\bfk^2)$  
the asymmetry is proportional to the combination of the intrinsic
heavy quark and heavy antiquark PDFs:
$\sqrt{Q_{\rm in}(x)}-\sqrt{\bar Q_{\rm in}(x)}$. 
When $Q_{\rm in}(x)$ and $\bar Q_{\rm in}(x)$ are degenerate,
one has ${\cal P}_{\rm as}(x) \equiv 0$. 
The ${\cal P}_{\rm as}(x)$ asymmetry is a new QCD asymmetry
arising due to interference of two Fock states in the proton
inducing intrinsic and extrinsic heavy quark content. 
Together with the 
$Q_{\rm in}(x) - \bar Q_{\rm in}(x)$ asymmetry considered in
Refs.~\cite{Sufian:2020coz,Brodsky:2022kef} it gives important
information about heavy quark distributions in the proton. 
Both asymmetries vanish at the limit $m_Q \to \infty$  
due to falloff of the normalization constants of the intrinsic and  
extrinsic contributions of heavy quark/antiquarks in the proton.  
As an application, we have calculated the asymmetry
${\cal P}_{\rm as}(x)$, which appear in the heavy quark structure function
$F_2^Q(x)$ and differential cross section $d\sigma_{e^-p}/dQ^2$ and
have found reasonable agreement with experimental results from
the H1 Collaboration~\cite{H1:2002voc}. 

\section*{Acknowledgments}

We thank Stan Brodsky for useful discussions. 
This work was funded by BMBF (Germany)
``Verbundprojekt 05P2021 (ErUM-FSP T01) -
Run 3 von ALICE am LHC: Perturbative Berechnungen
von Wirkungsquerschnitten f\"ur ALICE''
(F\"orderkennzeichen: 05P21VTCAA),
by ANID PIA/APOYO AFB180002 and AFB220004 (Chile),
by FONDECYT (Chile) under Grants No. 1191103, No. 1230160 and
No. 1230391 and by ANID$-$Millen\-nium
Program$-$ICN2019\_044 (Chile).

\end{document}